\documentclass[aps,jcp,twocolumn,showpacs,groupedaddress]{revtex4}
\usepackage{graphicx,amssymb,amsmath}
\begin{document}
\title{Minimal energy packings of nearly flexible polymers}
\author{Robert S. Hoy$^{1}$}
\email{rshoy@usf.edu}
\author{Jared Harwayne-Gidansky$^{2}$}
\author{Corey S. O'Hern$^{3,4,5,6}$}
\affiliation{$^{1}$Department of Physics, University of South Florida, Tampa, FL 33620}
\affiliation{Departments of Electrical Engineering$^2$, Mechanical Engineering \& Materials Science$^3$, Applied Physics$^4$, Physics$^5$, and $^6$Integrated Graduate Program in Physical and Engineering Biology, Yale University, New Haven, CT 06520}
\date{\today}
\begin{abstract}
We extend recent studies of the minimal energy packings of short
flexible polymers with hard-core-like repulsions and short-range
attractions to include bond-angle interactions with the aim of
describing the collapsed conformations of `colloidal' polymers.  We
find that flexible tangent sticky-hard-sphere (t-SHS) packings provide
a useful perturbative basis for analyzing polymer packings with
nonzero bending stiffness only for {\it small} ratios of the
stiffnesses for the bond-angle ($k_b$) and pair ($k_c$)  interactions, {\it i.e.}
$k_b^{\rm crit}/k_c \lesssim 0.01$ for $N<10$ monomers, and the
critical ratio decreases with $N$.  Below $k_b^{crit}$, angular
interactions give rise to an exponential (in $N$) increase in the
number of distinct angular energies arising from the diversity of
covalent backbone paths through t-SHS packings.  As $k_b$ increases
above $k_b^{crit}$, the low-lying energy landscape changes
dramatically as finite bending stiffness alters the structure of the
polymer packings.  This study lays the groundwork for
exact-enumeration studies of the collapsed states of t-SHS-like models
with larger bending stiffness.
\end{abstract}
\maketitle

\section{Introduction}
\label{sec:intro}

Protein folding and other examples of polymer collapse in dilute
solutions are complex processes that involve the cooperative motion of
thousands of atoms. A number of early studies of polymer and protein
structure and dynamics employed hard-sphere models with only steric
interactions, bond-length, and bond-angle constraints to understand
polymer elasticity\cite{flory59} as well as secondary structure\cite{ramachandran68} and packing in hydrophobic cavities\cite{richards77} in proteins. More recent studies have
implemented Monte Carlo, Brownian dynamics, and molecular dynamics
(MD) simulations that include excluded volume and stereochemical
constraints as well as solvent-mediated attractive interactions to
investigate numerous examples of polymer collapse.  However, recent
advances in colloidal synthesis and self assembly have enabled studies
of ``colloidal polymers'' with hard-core-like repulsions, short-range
attractions, and much greater chain flexibility than typical synthetic or
biological polymers\cite{sacanna10,solomon10}, which has prompted
renewed interest in simple polymer models.\cite{taylor03,foteinopoulou08,karayiannis09,laso09,taylor09,taylor09pre,seaton10,evans11,hoy12sm,ruzicka12}
We seek to develop the ability to theoretically predict which structures are most likely to
form for a given set of interactions and solvent conditions, and thus 
reducing the need for trial-and-error synthesis of self-assembled 
nanostructures.

The ground state packings of short, flexible tangent sticky hard
sphere (t-SHS) polymers have recently been characterized via complete
enumeration.\cite{hoy10} These packings possess a wider
range of symmetries and shapes than the ground state packings of
polymers with longer-ranged attractive (\textit{e.g.}\
Lennard-Jones\cite{schnabel09}) pair potentials.  Further, the number
$N_{\rm micro}$ of distinguishable, energetically degenerate
ground-state flexible t-SHS packings (``microstates'') grows
exponentially with the polymerization index $N$, with a corresponding
increase in the diversity of paths followed by the covalent backbones.

In this manuscript, we characterize the ground state packings for
short, {\it semi-flexible} t-SHS polymers with finite bending
stiffness $k_b$.  In particular, we calculate the potential energy of
semi-flexible polymer packings as a function of $k_b$ and identify the
maximum bending stiffness $k_b^{\rm crit}$ below which the reference
set of ground-state flexible t-SHS packings are the lowest energy
structures.  We also examine the structural differences between the
lowest energy semi-flexible structures and the completely flexible
reference set that appear as $k_b$ approaches and exceeds
$k_b^{crit}$.  The crossover occurs in a range of $k_b$ lying between
estimates of $k_b^{crit}$ from a simple analytic criterion and results
from MD simulations of polymer collapse.  In this range of $k_b$, the
low-lying energy minima in the landscape change dramatically.  In
general, states with $k_b > k_b^{crit}$ are less collapsed than flexible t-SHS
packings since it is energetically unfavorable for such chains to form
the maximum number of pair contacts allowed by steric constraints.
Our results will inform future exact-enumeration studies of polymer
packings with larger bending stiffness.

\section{Model and Methods}
\label{sec:modelmethods}

\begin{figure}[htbp]
\includegraphics[width=3.375in]{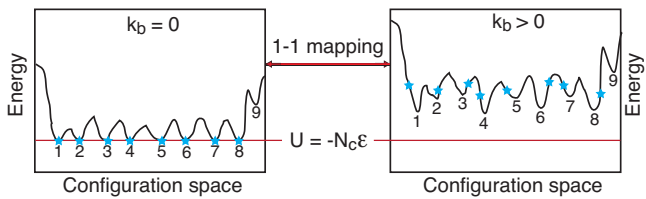}
\caption{Schematic of the low-lying minima in the energy landscape for
polymer packings as a function of the configurational degrees of
freedom $\{{\vec r}\}$.  For vanishing bending stiffness $k_b = 0$
(left), the ground-state flexible polymer packings (stars) are
energetically degenerate and correspond to the lowest $N_{\rm micro}$
inherent structures (minima of the black curve).  Increasing $k_b$
(right) deforms the energy landscape and raises the energies of the
$N_{\rm micro}$ minima that correspond to the flexible polymer
packings.  For $k_b < k_b^{\rm crit}$, the $N_{\rm micro}$ lowest
energy semi-flexible polymer packings correspond to the $N_{\rm
micro}$ ground-state flexible polymer packings.}
\label{fig:mappingassump}
\end{figure}

\begin{figure}[htbp]
\includegraphics[width=3in]{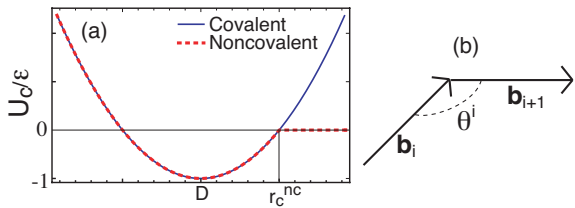}
\caption{Schematic of (a) the short-range attractive and harmonic
bond-length potentials $U_c(r)$ (Eq.\ \ref{eq:perturbed}) and (b) the
definition of the bond angle $\theta^i$ for the harmonic bond-angle
potential $U_{\rm bend}^i$ (Eq.\ \ref{eq:ubend}).}
\label{fig:potential}
\end{figure}

We consider a polymer model with harmonic pair and bond-angle
potentials with respective ``spring constants'' $k_c$ and $k_b$, and
equilibrium bond angle $\theta_{eq}$ (Fig.\ \ref{fig:potential} and
Eqs.\ \ref{eq:perturbed}-\ref{eq:ubend}).  The ground-state flexible
($k_b=0$) t-SHS polymer packings can provide a useful reference set of
configurations for studying the lowest energy semi-flexible ($k_b>0$)
t-SHS polymer packings when there is a one-to-one mapping from the
$N_{\rm micro}$ ground state polymer packings for $k_b=0$ to the
$N_{\rm micro}$ lowest energy polymer packings with finite bending
stiffness.  In this case, the $N_{\rm micro}$ zero-$k_b$ t-SHS
packings lie within the basins\cite{stillinger95} of the $N_{\rm
micro}$ lowest-energy packings at finite $k_b$ as shown in
Fig.~\ref{fig:mappingassump} (right).  Increasing $k_{b} >0$ deforms
the energy landscape, raises the energies of the $N_{\rm micro}$
flexible polymer packings, and breaks their degeneracy.  For $k_b <
k_b^{\rm crit}$, the $N_{\rm micro}$ lowest energy semi-flexible
polymer packings correspond to the $N_{\rm micro}$ ground-state
flexible polymer packings.  In contrast, for $k_b > k_b^{\rm crit}$
the energy landscape is sufficiently deformed that new low-energy
packings (that do not belong to the set of $k_b=0$ ground-state
polymer packings) are among the $N_{micro}$ lowest.  For example, in
Fig.~\ref{fig:mappingassump} with $N_{\rm micro}=8$, if the energy of
polymer packing $9$ drops below any of the energies of packings $1$
through $8$, there is no one-to-one correspondence between the $N_{micro}$ lowest-energy microstates and $k_b > k_b^{\rm crit}$.

\begin{table}[htbp]
\caption{Number of microstates $N_{\rm micro}$ for maximally contacting
flexible t-SHS packings, where $N_{\rm micro}$ for $N > 6$ includes a
factor of two associated with packings possessing chiral
enantiomers.\cite{arkus09}  The number of contacts $N_c$ is the number of $\{i,j\}$ pairs with $r_{ij} < r_c$.}
\begin{ruledtabular}
\begin{tabular}{lcc} 
$N$ & $N_c$ & $N_{\rm micro}$\\
6 & 12 & 50\\
7 & 15 & 700\\
8 & 18 & 6429\\
9 & 21 & 122060\\
\end{tabular}
\end{ruledtabular}
\label{tab:MNtab}
\end{table}

All of the $N_{\rm micro}$ distinguishable polymeric paths
(microstates) through $N$-monomer, $N_c$-contact flexible t-SHS
packings are obtained via the complete enumeration procedure described
in Ref.\ \cite{hoy10}.  Values of $N_{\rm micro}$ for ground-state
(maximal-$N_c$) flexible t-SHS packings of $6 < N < 9$ monomers are given in
Table \ref{tab:MNtab}.  

Many computational models for polymers coarse-grain at
the level of several monomers per coarse-grained spherical bead.
However, since colloidal polymers are reasonably modeled by spheres
with hard-core-like repulsions and short-range interactions, we
coarse-grain at the level of one sticky sphere per colloidal monomer.
Thus our range of $N$ is appropriate for modeling experiments like
those in Refs.\ \cite{sacanna10,solomon10}, which studied colloidal polymers
with $N \leq 8$.

The pair potential for t-SHS polymers with spherical monomers of diameter $D$
is\cite{yuste93,hoy10}:
\begin{equation}
U_{ss}(r) = \Bigg{\{}\begin{array}{ccccr}
\infty & , & r < D & &\\
-\epsilon & , & r = D & &\\
0 & , & r > D & , & \textrm{noncovalent}\\
\infty & , & r > D & , & \textrm{covalent},
\end{array}
\label{eq:stickyspherepot}
\end{equation}
where $r$ is the center-to-center separation between monomers and
$-\epsilon$ is the contact energy.  In Eq.\ \ref{eq:stickyspherepot},
the term ``covalent'' (``noncovalent'') refers to monomers that are
(not) chemically adjacent and permanently connected.  
For numerical calculations, we employ a continuous version of
$U_{ss}(r)$ with short-range attractive interactions and harmonic
bond-length constraints illustrated in Fig.~\ref{fig:potential}(a),
\begin{equation} 
U_{c}(r) = \bigg{\{}\begin{array}{ccc}
-\epsilon + \displaystyle\frac{k_c}{2}\left(\frac{r}{D}-1\right)^{2} & , & r < r_c\\
& &\\
0 & , & r \ge r_c,
\end{array}
\label{eq:perturbed}
\end{equation}
where $k_c$ is in units of $\epsilon$, $r_c/D = \infty$ for covalently bonded monomers and $r_c = r_c^{nc} = D(1 + \sqrt{2/k_c})$ for noncovalently bonded monomers.  Note that
$U_{ss}(r)$ is the $k_c \to \infty$ limit of $U_c$ in
Eq.~\ref{eq:perturbed}.

To model finite bending stiffness, we employ
a harmonic bond-angle potential used in many previous
computational studies of organic molecules, peptides, and
proteins~\cite{cross33,toxvaerd97}:
\begin{equation}
U^i_{\rm bend} = \displaystyle\frac{k_b}{2}\displaystyle\frac{(\theta^i-\theta^i_{eq})^2}{\pi^2}.
\label{eq:ubend}
\end{equation}
The bond angle $\theta^i$ between adjacent monomers ($i$, $i+1$, $i+2$) is
defined as $\theta^i = \pi -
\cos^{-1}[(\vec{b}_{i}\cdot\vec{b}_{i+1})/|b_i b_{i+1}|]$, with
$\vec{b}_{i} = \vec{r}_{i+1}-\vec{r}_{i}$.  $U^i_{\rm bend}$ is minimized at
the equilibrium bond angle $\theta^i = \theta^i_{eq}$, and $k_b$ is
expressed in units of $\epsilon$.

The total potential energy of a $N$-monomer chain is 
\begin{equation}
\begin{array}{lll}
U_{\rm chain} & = & U_c^{\rm tot} + U_b^{\rm tot} \nonumber \\
& = & \sum_{i=1}^{N-1} \sum_{j>i}^{N} U_{c}(r_{ij}) +
\sum_{i=1}^{N-2} U_{\rm bend}^{i}.
\end{array}
\label{eq:uangchain}
\end{equation}
For $k_b = 0$ and the limit $k_c \to \infty$, configurations
corresponding to local minima of $U_{\rm chain}$ are clearly identical to
those obtained in studies of flexible t-SHS packings with no bond-angle
interactions, \textit{i.e.}\ all $N$-monomer, $N_c$-contact microstates
have $U_{\rm chain} = -N_c\epsilon$.\cite{footkc3200} 

Bond-angle interactions break this degeneracy; $U_{\rm chain}$
increases by different amounts for each microstate.  For t-SHS
packings, to leading order (LO), the increase in energy $U_{\rm
chain}(k_b) - U_{\rm chain}(k_b=0)$ scales linearly with $k_b$.  To
see why this is so, note that for $k_c\to\infty$, increasing the
bending stiffness does not change the structure of the polymer
packings. As $k_c$ becomes very large, the energetic penalty for
changing the intermonomer distances $r_{ij}$ for all pairs $\{i,j\}$
becomes correspondingly large. In this limit, changing any of the bond
angles alters the intermonomer distances and necessarily implies
contact-breaking ($r_{ij} > D$) or intermonomer overlap ($r_{ij} <
D$).\cite{hoy10}

Each ``reference'' $k_b = 0$ microstate
possesses a set of $N-2$ reference bond angles $\{\theta^{ref}_1,\
\theta^{ref}_2,\ ...,\ \theta^{ref}_{N-2}\}$.  
The LO prediction for the energy of the \textit{j}th microstate is
\begin{equation}
U_{\rm chain,LO}^{j} = -N_c \epsilon + \displaystyle\frac{k_b}{2\pi^2} c_j(N,\theta_{eq})
\label{eq:LOUchaink2}
\end{equation}
where 
\begin{equation}
c_j(N,\theta_{eq}) \equiv  \sum_{i=1}^{N-2} (\theta^{ref,j}_i - \theta_{eq})^2
\label{eq:cjtheta0}
\end{equation}
and we assume $\theta_{eq}^i = \theta_{eq}$ for all $i$.  Since all
$c_j$ are available from the flexible t-SHS packings, LO predictions
for the energy of polymer packings for arbitrary $k_b$ and
$\theta_{eq}$ can be made using only information from the flexible
reference polymer packings.  Next-to-leading order (NLO) corrections
to $U_{\rm chain}$ are negative and are expected to scale as $k_b/k_c$.
Below, we will analytically and numerically calculate the
total energy $U_{\rm chain}$ for the reference packings versus $k_b$ and
determine $k_b^{\rm crit}$ as a function of $N$ and $\theta_{eq}$.

\section{Results}
\label{sec:results1}

\begin{figure}[htbp]
\includegraphics[width=3in]{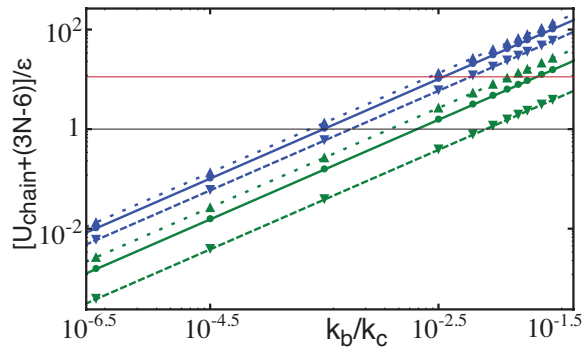}
\caption{Comparison of the leading-order expressions for the energy
$U_{\rm chain}$ of $N=8$ semiflexible polymer packings in
Eq.~\ref{eq:LOUchaink2} (dashed, $U_{\rm min}$; solid, $U_{\rm avg}$;
dotted, $U_{\rm max}$) to the numerical results for the energy
of minimized t-SHS packings in the presence of bond-angle interactions
(downward triangles, $U_{\rm min}$; circles, $U_{\rm avg}$; upward
triangles, $U_{\rm max}$) as a function of $k_b/k_c$.  Results for
$\theta_{eq} = \theta_{tet}$ and $\pi$ are shown in green and blue,
respectively.  $U_{\rm chain}/\epsilon = -(N_c-1)$ and $U_{\rm chain} = 0$ are respectively indicated by the black and red horizontal lines.}
\label{fig:analytvsnumeric}
\end{figure}

Predictions for the minimum, maximum, and average
energies of polymer packings ($U_{min},\ U_{max},\ U_{avg}$) from
Eq.~\ref{eq:LOUchaink2} for $N=8$ and two physically relevant
$\theta_{eq}$ are shown in Fig.~\ref{fig:analytvsnumeric}. 
Polymers with tetrahedral bond-angle order
have $\theta_{eq} \simeq \theta_{tet}
=\cos^{-1}(-1/3) = 109.47^{\circ}$, while $\theta_{eq} = \pi$
corresponds to polymers with linear bond-angle
order.  The bond-angle energies are higher for $\theta_{eq} = \pi$
since the compact structure of small-$N$ flexible t-SHS packings allows
few ``straight trimers'' (angles with $\theta^i = \pi$).
To validate the leading-order expressions for the energy of
semiflexible polymer packings in Eq.~\ref{eq:LOUchaink2}, we employed
energy minimization techniques (described in the Appendix) to relax
the t-SHS configurations in the presence of bond-angle
interactions. The LO expressions for the minimum $U_{\rm
min}$, average $U_{\rm avg}$, and maximum $U_{\rm max}$ chain energies
closely agree with the numerical results over a wide range of
$k_b/k_c$ as shown in Fig.~\ref{fig:analytvsnumeric}.  
Both the numerical results and LO expressions 
for $U_{\rm chain}$ are nearly linear in
$k_b/k_c$ over the full range studied.  

\begin{figure}[htbp]
\includegraphics[width=3in]{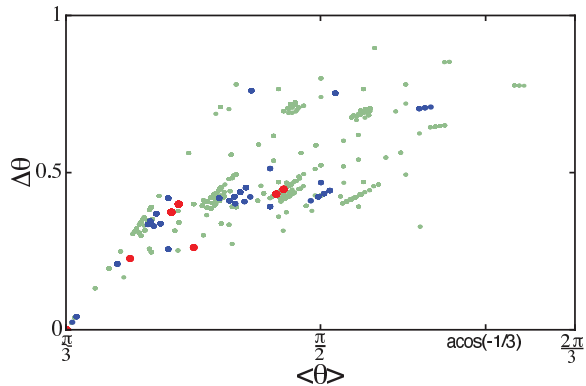}
\caption{Bond-angle dispersion for flexible t-SHS polymer packings.
Each data point represents the mean bond angle $\langle \theta
\rangle$ and the root-mean-square deviation $\Delta \theta$ for a
single flexible t-SHS polymer packing.  Large red, medium blue, and
small green circles represent data for $N = 6.\ 7,\ {\rm and}$ $8$,
respectively.}
\label{fig:thetascatter}
\end{figure}

Note that for $\theta_{eq} = \theta_{tet}$, $(U_{\rm max} +
N_c\epsilon)/(U_{\rm min} + N_c\epsilon) \simeq 7$ because there are
large bond angle fluctuations $\Delta \theta$ in the flexible t-SHS
packings\cite{hoy10,arkus09,hoy12pre} as shown in
Fig. \ref{fig:thetascatter}.  The variation in $\langle \theta
\rangle$ and $\Delta \theta$ increases rapidly with $N$
(Fig.~\ref{fig:thetascatter}) and suggests that $k_b^{crit} \ll 1$.

To visually estimate $k_b^{crit}$ from Fig.\
\ref{fig:analytvsnumeric}, we note that freely-rotating chain
conformations with no pair contacts and $\theta^i=\theta_{eq}$ for all
bond angles have $U_{\rm chain}=-(N-1)\epsilon$.  Thus it is clear
that $k_b/k_c > k_b^{\rm crit}/k_c$ for $U_{\rm chain} > 0$, and that
the critical bending stiffness $k_b^{\rm crit}$ 
is below $k_b^{\rm crit}/k_c \sim 10^{-1.5}$.
A more refined (if heuristic) estimate of $k_b^{\rm crit}$ can be obtained as
follows.  The $N_{\rm micro}(N)$ flexible t-SHS packings all possess
$N_c(N)$ pair contacts (Table \ref{tab:MNtab}) and energy
$-N_c(N)\epsilon$.  Any polymer packing with fewer than $N_c(N)$ pair
contacts must have $U_{\rm chain} \geq -(N_c -1)\epsilon$ since the bending
energy is strictly positive.  Conversely, any t-SHS polymer packing with
$N_c$ pair contacts must correspond (in the sense of Fig.\
\ref{fig:mappingassump}) to one of the $N_{\rm micro}(N)$ flexible t-SHS
packings.  Therefore $k_b^{\rm crit}(N,\theta_{eq})$ is set by the largest
$c_j(N,\theta_{eq})$ (Eq.\ \ref{eq:cjtheta0}) for these packings and
the condition that the reduced bending energy is less than unity,
which gives 
\begin{equation}
k_b^{\rm crit}(N,\theta_{eq}) \le 2\pi^2/c_j^{max}(N,\theta_{eq}).
\label{eq:kbcriteq1}
\end{equation}

\begin{figure}[htbp]
\includegraphics[width=3in]{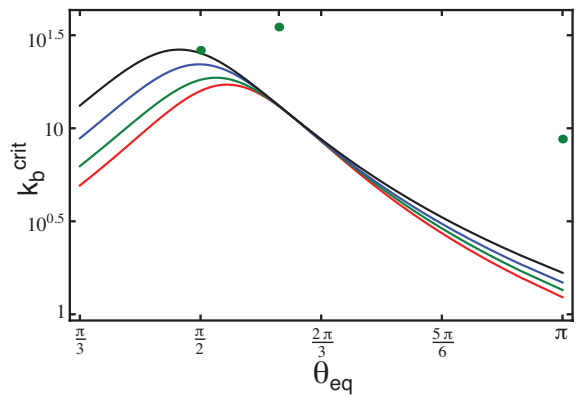}
\caption{Estimates of $k_b^{\rm crit}(N,\theta_{eq})$ from
Eq.~\ref{eq:kbcriteq1} for $N= 6$ (black line), $7$ (blue line), $8$
(green line), and $9$ (red line) and from MD simulations of finite-$k_b$ 
t-SHS polymers with $N=8$ (green circles).}
\label{fig:kbcritaest}
\end{figure}

As shown in Fig.~\ref{fig:kbcritaest}, $k_b^{\rm crit}$ decreases
monotonically with increasing $N$ for all $\theta_{eq} \lesssim \pi/2$
and $\theta_{eq} \gtrsim 2\pi/3$, which stems from the increasing
angular dispersion with $N$ shown in Fig.~\ref{fig:thetascatter}.  In
addition, for increasing $\theta_{eq}$, $k_b^{\rm crit}$ reaches a
peak at $\theta_{\rm max}$ that increases with $N$, and then decreases
for $\theta_{eq} > \theta_{\rm max}$.  For the $N$ considered
here, the peak occurs at a rather small $\theta_{\rm max} \lesssim
\pi/2$ because of the compactness of the flexible t-SHS reference
polymer packings.  We expect the decrease in $k_b^{crit}$ with
increasing $N$ to continue for $N>9$.

To check the analytic prediction of Eq.~\ref{eq:kbcriteq1} for
$k_b^{\rm crit}$, we performed MD simulations of semiflexible t-SHS
chain collapse for $N=8$.  Monomers interact via the pair and
bond-angle potentials in Eqs.\ \ref{eq:perturbed} and~\ref{eq:ubend},
with $k_c = 3200$ and a range of $\theta_{eq}$ and $k_b$.  For each
$\theta_{eq}$, $k_b$, and $k_c$, $N_{\rm S}=1000$ systems with
independent self-avoiding random walk initial polymer configurations
were slowly quenched from high temperature ($k_B T_{i} = 4\epsilon$)
to $k_B T_f < 10^{-10}\epsilon$ using an isokinetic
thermostat~\cite{brown} and quench protocol $T =
T_{i}\exp{[-t/(10^{3}\tau)]}$, where $\tau = \sqrt{mD^{2}/\epsilon}$
is the unit of time and $m$ is the monomer mass.\cite{footMD2} An
overestimate of $k_b^{\rm crit}$ can be obtained by measuring the
lowest $k_b$ at which the minimum energy collapsed states obtained via
MD have chain energies below $U_{\rm min}$; these cannot correspond to
any of the $N_{\rm micro}$ flexible t-SHS inherent structures.  Note
that in the sticky-hard-sphere limit, $k_b^{crit}$ is set by the
structure of t-SHS packings ({\it i.e.} the number $N_c(N)$ and types
of inter-monomer contacts).

A comparison of $k_b^{\rm crit}$ from Eq.~\ref{eq:kbcriteq1} and
estimates from the MD simulations is shown in
Fig.~\ref{fig:kbcritaest} for several $\theta_{eq}$.  The green
circles illustrate the lowest bending stiffnesses $k_b^*$ at which
polymer packings with energies lower than $U_{\rm min}$ are formed.
In all cases, the polymer packings from MD simulations with energies
$U_{\rm chain} < U_{\rm min}$ possess 2-3 fewer pair contacts than the
flexible t-SHS inherent structures and have $k_b^*$ such that
$[U_{\rm max}(k_b^{*})+(3N-6)]\simeq (2-3)\epsilon$.  Thus,
energetically favorable rearrangements of t-SHS polymer packings due
to finite bending stiffness typically require breaking two or three pair
contacts.

\begin{figure}[htbp]
\includegraphics[width=3.0in]{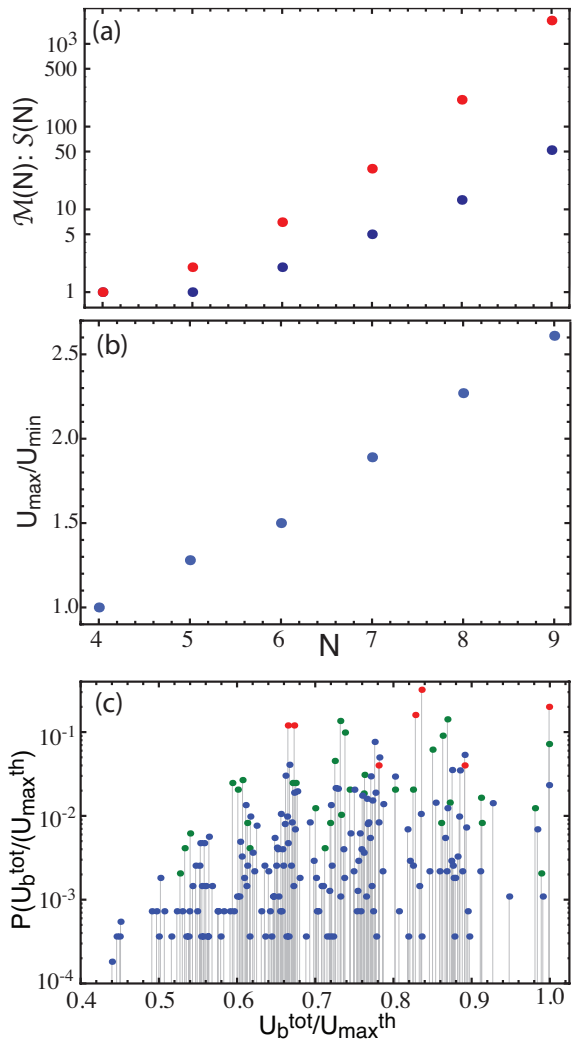}
\caption{(a) Number of nonisomorphic packings (macrostates)
$\mathcal{M}(N)$\cite{hoy10} (blue circles) and non-degenerate angular
energy levels $\mathcal{S}(N)$ (red circles), and (b) ratio $U_{\rm
max}/U_{\rm min}$ plotted versus $N$ for $\theta_{eq} = \pi$, in
the limit $k_b \to 0$. (c) The distribution of angular energies
$P(U_b^{\rm tot}/U_{\rm max}^{th})$ for $N = 6$ (red) $7$ (green), and
$8$ (blue).}
\label{fig:degensplittingpi}
\end{figure}

We now return to the picture of Fig.~\ref{fig:mappingassump} and
analyze how the introduction of nonzero bending stiffness alters the
low-lying energy landscapes of t-SHS polymers.  Figure
\ref{fig:degensplittingpi}(a) compares the variation of the
structural and angular energy degeneracy splittings for $4 \leq N \leq
9$ and $\theta_{eq} = \pi$ in the limit $k_b \to 0$.  The reference
t-SHS packings can be classified into $\mathcal{M}(N)$ distinguishable
(structurally nonisomorphic\cite{hoy10}) ``macrostates'' possessing
different shapes, {\it i.e.} sets of squared interparticle distances
$\{r_{ij}^2\}$)\cite{hoy10}, and $\mathcal{M}(N)$ grows exponentially
with $N$.  In general, the different polymeric paths through any given
macrostate have many different combinations of angles (Fig.\
\ref{fig:thetascatter}), and thus, at finite $k_b$, different
energies.  The number $\mathcal{S}(N)$ of distinct angular energy
levels $U_b^{\rm tot}$ for any given $\theta_{eq}$ also increases
exponentially with $N$ and scales roughly as $N_{\rm
micro}/\mathcal{M}$.  Figure \ref{fig:degensplittingpi}(b)
illustrates the range of energy levels, $R = U_{\rm max}/U_{\rm min}$,
for $N=8$ t-SHS polymers in the $k_b \rightarrow 0$ limit.  In
contrast to the exponential increase in the number of distinct energy
levels, $R$ increases only (roughly) linearly with $N$.  This can be
understood as follows. Since $U_{\rm max}$ has an upper bound
\begin{equation}
U_{\rm max}^{th} = (N-2)(k_b/2\pi)(2\pi/3)^3
\label{eq:umaxth}
\end{equation}
imposed by steric constraints, the spacing between energy levels
decreases as $\mathcal{S}$ increases.  This is illustrated in Figure
\ref{fig:degensplittingpi}(c), which presents results for the
distribution $P(U_b^{\rm tot}/U_{\rm max}^{th})$ of the total angular
energies relative to $U_{\rm max}^{th}$ for $6 < N < 8$. Results for
all $N$ show $\delta$-function-like peaks, and the spacing between
energy levels decreases as $N$ increases. The distribution $P(U_b^{\rm
tot}/U_{max}^{th})$ also broadens as $N$ increases.  We have also
examined $P(U_b^{tot}/U_{max}^{th})$ for $\theta_{eq} = \theta_{tet}$. 
While the distribution shifts to lower values of $U_b^{\rm tot}$
since $\theta_{tet}$ is a commonly occurring angle in t-SHS packings,
qualitative trends with $N$ are similar.  Note that Fig.\
\ref{fig:degensplittingpi} shows analytic results calculated from the
values of $c_j(N,\theta_{eq})$.  At finite $k_b < k_b^{\rm crit}$, numerical results for $P(U_b^{\rm
tot}/U_{max}^{th})$ show that packings with the same $\{\theta^i\}$ but topologically distinct orderings of the $\theta^i$ relax differently towards $\theta_{eq}$, and the distribution of angular energies becomes more continuous.
 
\begin{figure}[htbp]
\includegraphics[width=3.0in]{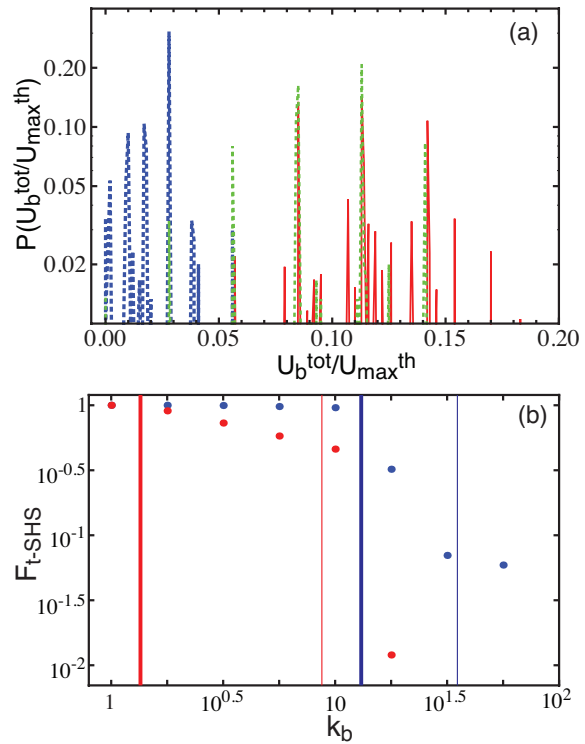}
\caption{(a) Distributions $P(U_{b}^{\rm tot}/U_{\rm max}^{th})$ of
the total angular energy $U_b^{\rm tot}$ relative to $U_{\rm
max}^{th}$ in the $k_b \to 0$ limit (red), $k_b = 1$ (green) and
$10^{1.5}$ (blue) for $N=8$ and $\theta_{eq} = \theta_{tet}$.  (b)
Fraction $F_{t-SHS}$ of collapsed packings corresponding to t-SHS
microstates as a function of $k_b$ for $\theta_{eq} = \pi$ (red
circles) and $\theta_{tet}$ (blue circles). The heavy and light solid
lines correspond to estimates of $k_b^{crit}$ from Eq.\
\ref{eq:kbcriteq1} and MD estimates of $k_b^{*}$ (Fig.\
\ref{fig:kbcritaest}), respectively.}
\label{fig:finitekb}
\end{figure}

Next we examine features of the low-lying energy landscape for
packings with $k_b \simeq k_{b}^{crit}$.  Figure \ref{fig:finitekb}(a)
contrasts the probability distributions $P(U_b^{\rm tot}/U_{max}^{th})$
for systems with $N = 8$ and $\theta_{eq}=\theta_{tet}$ in the limit
$k_b\to0$, for two finite values of $k_b$: $k_b=1 \ll k_b^{crit}$,
and $k_b = 10^{1.5} \simeq k_b^{*}$.  Results for the larger two $k_b$
are obtained using the same MD protocol described above, while results
for $k_b\to0$ are evaluated analytically using 
$c_j(N,\theta_{eq})$.  Results for $k_b \to 0$ and $k_b=1$ are
essentially equivalent in that all packings correspond to t-SHS
packings.  Small differences arise from the MD quench protocol, which
favors formation of packings with lower $U_b^{\rm tot}$.

Our results from the MD simulations for $k_b \simeq k_b^{*}$
illustrate that significant changes occur in the low-lying energy
landscape for semi-flexible t-SHS polymers.  The distribution
$P(U_{b}^{\rm tot}/U_{\rm max}^{th})$ overlaps those for $k_b\to0$
only at the lowest peak ($U_{b}^{\rm tot}/U_{\rm max}^{th} \simeq
.058$).  The remainder of the collapsed states have lower $U_{b}^{\rm
tot}/U_{\rm max}^{th}$ and do not correspond to flexible t-SHS
packings.  These states possess fewer pair contacts ($N_c < 18$) and
hence greater freedom to reduce $U_b^{\rm tot}$ by relaxing bond
angles toward $\theta=\theta_{eq}$.  Figure \ref{fig:finitekb}(b)
shows the fraction of collapsed MD-quenched configurations that form
ground-state flexible t-SHS polymer packings as a function of $k_b$.
We find that the low-lying energy landscape for t-SHS polymers changes
dramatically in the $k_b^{crit} < k_b \lesssim k_b^{*}$ regime when the 
structural properties of the lowest $N_{\rm micro}$ energy minima
become significantly different from flexible t-SHS packings.

\section{Conclusions}
\label{sec:conclude}

In this manuscript, we characterized tangent-sticky-hard-sphere polymer
packings with finite bending stiffness.  We have shown that t-SHS
polymers possess the same low-lying energy landscapes for $k_b <
k_b^{crit}$, where $k_{b}^{crit}$ depends strongly on the equilibrium
bond angle $\theta_{eq}$ and decreases with increasing degree of
polymerization $N$.  Angular interactions introduce new energy levels 
(compared to flexible t-SHS packings), whose number increases
exponentially with $N$.  As $k_b$ increases above $k_{b}^{crit}$, the t-SHS
energy landscape breaks down as a useful reference
for the low-lying energy-landscape of compact finite-stiffness polymers,
revealing an interesting regime in which minimizing pair and
bond-angle energies compete.  In future studies, we will enumerate and
characterize the structural and mechanical properties of minimal
energy semiflexible polymer packings with $k_b > k_b^{crit}$ using
advanced sampling techniques.

Many recent theoretical studies have examined the
structure of collapsed, flexible polymer chains (\textit{e.g.}\ Refs.\
\cite{taylor03,foteinopoulou08,karayiannis09,laso09,taylor09,taylor09pre,seaton10,evans11,hoy12sm,ruzicka12}).
It would be interesting to check how these are affected by a small but
finite bending stiffness.  For example, even a small $k_b$, since it
would alter the local ground-state structure of the polymer packings,
seems likely to produce a strong effect on the jamming transition of
densely packed collapsed polymers, which for fully flexible chains
occurs \cite{foteinopoulou08,karayiannis09,laso09} at random close
packing\cite{torquato00} as for monomers.

Support from NSF Award No.\ DMR-1006537 is gratefully acknowledged.
This work also benefited from the facilities and staff of the Yale
University Faculty of Arts and Sciences High Performance Computing
Center and NSF grant No.\ CNS-0821132 that partially funded
acquisition of the computational facilities.

\begin{appendix}

\section{Structure solver for packings with finite $k_c$ and $k_b$}

\begin{figure*}[htbp]
\includegraphics[width=6.8in]{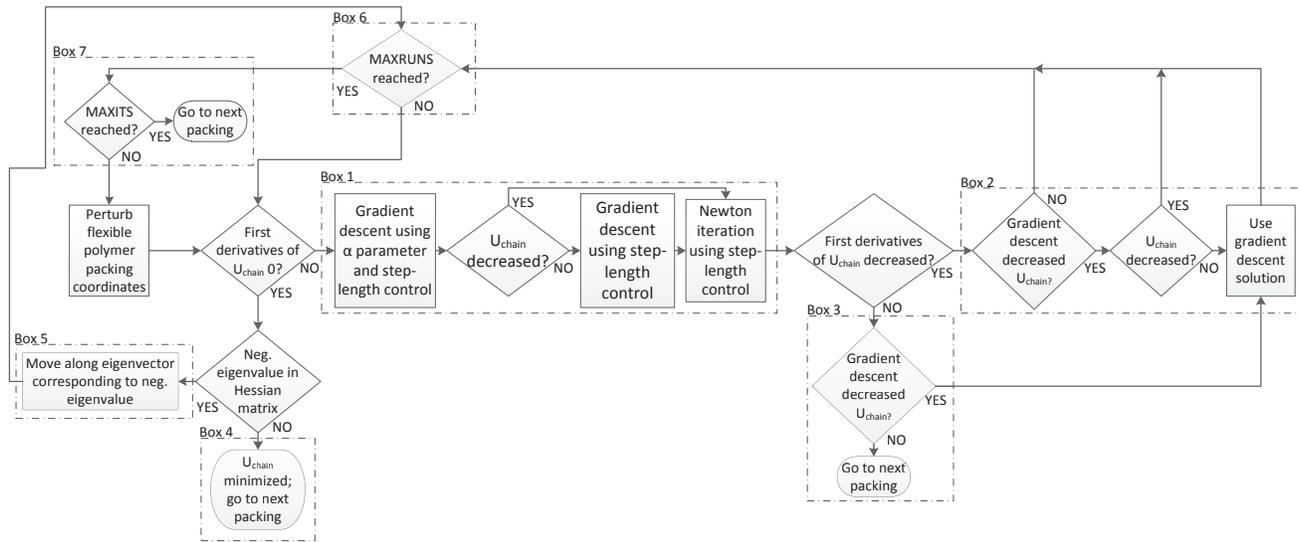}
\caption{Flowchart depicting the algorithm used to find the
mechanically-stable, minimum-energy configuration $\{\vec{r}_{min}\}$
closest to each flexible packing configuration $\{\vec{r}_{flex}\}$.}
\label{fig:flowchart}
\end{figure*}

We find the mechanically stable, minimal energy configuration
$\{\vec{r}_{min}\}$ closest to each flexible packing
$\{\vec{r}_{flex}\}$ by starting from configurations
$\{\vec{r}_{pert}\}$ representing small random
perturbations\cite{footperturb} away from $\{\vec{r}_{flex}\}$, and
minimizing $U_{\rm chain}$. Energy minimization is carried out using the
``dog-leg'' method \cite{dennis96}, which is a combination of gradient
descent and Newton's method \cite{press07}.  Our numerical algorithm
is schematically depicted in Figure \ref{fig:flowchart}.

For finite $k_b$, the first spatial derivatives of $U_{\rm chain}$
evaluated at $\{\vec{r}\} \neq \{\vec{r}_{min}\}$ are generally
nonzero.  We approach $\{\vec{r}\} = \{\vec{r}_{min}\}$ using gradient
descent and a second-order expansion of $f(\{\vec{r}\}) \equiv U_{\rm
chain}(\{\vec{r}\})$.  This descent iteratively reduces $f$ using
\begin{equation}
f_{n}^*=f_n- \alpha \nabla f_n,
\label{eq:dogleg1}
\end{equation}
where $f_n$ is the value of $f$ at the nth minimization step,
$f^*_n$ is $U_{\rm chain}$ at the ``knee'' of the dog-leg
\cite{footaftergrad}, the optimal step length is \cite{dogleg}
\begin{equation}
\alpha=\displaystyle\frac{\nabla f  (\nabla f)^T} {\nabla f H(f) (\nabla f)^T},
\label{eq:dogleg2}
\end{equation}
and $H(f_n)$ is the $3N\times 3N$ Hessian matrix of $f_n$.  After the
gradient descent step, we perform a least-squares Newton iteration on
$\nabla f_n$ using step-length control \cite{press07} (Box 1):
\begin{equation}
f_{n+1}=f^*_n- \left( H^T(f^*_n) H(f^*_n) \right)^{-1} H^T(f^*_n)  \nabla f^*_n.
\label{eq:dogleg4}
\end{equation}
If the Frobenius norm $||\nabla f_{n+1}||_2$ of $\nabla f_{n+1}$
decreases (\textit{i.e.} $||\nabla f_{n+1}||_2 < ||\nabla
f^*_{n}||_2$), we check whether$f_{n+1} < f^*_n$. If $f_{n+1} >
f^*_n$, we set $f_{n+1}=f^*_{n}$; if $f_n^{*} \geq f_n$, the gradient
descent has failed and we set $f_n^{*} = f_n$ in Eq.\
\ref{eq:dogleg4}. We then check whether $||\nabla f_{n+1}||_2 \le
||\nabla f^*_{n}||_2$; if so, Equations
\ref{eq:dogleg1}-\ref{eq:dogleg4} represent one complete minimization
step taking $f_n \to f_{n+1}$ (Boxes 1-2).  Otherwise, if $||\nabla
f_{n+1}||_2 > ||\nabla f^*_{n}||_2$ the minimization algorithm has
failed (Box 3); however, this occurs only for $k_b$ much larger than
those considered here.

The minimization procedure reduces the first spatial derivatives of
$U_{\rm chain}$ to zero. When this condition is satisfied ($\nabla f =
0$), we check whether $\{\vec{r}\}$ represents a mechanically stable
solution, \textit{i.e.} whether all $3N-6$ nontrivial eigenvalues of
$H$ are positive (Box 4). Saddle-point solutions are avoided (when
encountered) by perturbing coordinates along a negative eigenvector of
the Hessian matrix (Box 5).

Should the solver reach the maximum number of iterations for
a particular $\{\vec{r}_{pert}\}$ (MAXRUNS; Box 6) we attempt to minimize
a different $\{\vec{r}_{pert}\}$, for up to MAXITS different
$\{\vec{r}_{pert}\}$ (Box 7).  We found that $\rm
MAXITS=1000$ different $\{\vec{r}_{pert}\}$ with a maximum of $50$
iterations per $\{\vec{r}_{pert}\}$ to be sufficient to minimize the
energy or determine that no minimum
that preserves the adjacency matrix exists for the $N \leq 9$
polymer packings considered here.

Finally, note that the time required for our structure solver to find energy minima is insignificant for a given $k_b$ and $\theta_{eq}$.  However, two factors limit the present study to $N=9$: (a) The time required to enumerate the flexible perturbative basis increases faster than exponentially with $N$, as reported in Refs.\ \cite{hoy10,hoy12pre}.  (b) While flexible polymer packings have been generated for $N$ up to $11$, the ground state packings for $N = 10\ \rm{and}\ 11$ show much less structural diversity (e.g.\ fewer possible $\theta^i$) and hence are less suitable for perturbative studies of the effect of angular stiffness.

\end{appendix}


\begin{thebibliography}{32}
\bibitem[{foo({\natexlab{a}})}]{footkc3200}
\bibinfo{note}{We employ a large value of $k_c$ ($3200$), which because of its
  small $r_c$, is in the SHS limit and allows a one-to-one mapping between
  flexible t-SHS polymer packings and degenerate ground states in the $k_b \to
  0$ limit. For softer interactions with larger values of $r_c$,
  second-nearest-neighbor pair interactions would break this degeneracy.
  However, an analysis similar to that presented below but taking account of
  this degeneracy-breaking could be performed.}
\bibitem[{foo({\natexlab{b}})}]{footMD2}
\bibinfo{note}{The quench rate employed is close to the slow-quench limit for
  the $k_c = 3200$.}
\bibitem[{foo({\natexlab{c}})}]{footperturb}
\bibinfo{note}{Each monomer position is perturbed by a maximum amplitude
  $|\delta \vec{r}|_{\rm max} = 0.2$, which was chosen to optimize the
  efficiency of the structure solver while satisfying the requirement $|\delta
  \vec{r}|_{max} < (\sqrt{2}-1)/2$ for preserving chain topology.}
\bibitem[{foo({\natexlab{d}})}]{footaftergrad}
\bibinfo{note}{In the event that $U_{\rm chain}$ does not decrease, we attempt
  gradient descent with step-length control without the $\alpha$ parameter,
  \textit{i.e.} $f_{n}^*=f_n-\nabla f_n$.}
\end{thebibliography}
\end{document}